# The enduring mystery of the solar corona

Physicists have long known that the Sun's magnetic fields make its corona much hotter than the surface of the star itself. But how – and why – those fields transport and deposit their energy is still a mystery, as **Philip G Judge** explains

**Philip G Judge** is senior researcher in physics at the High Altitude Observatory in Boulder, Colorado, US, e-mail judge@ucar.edu

Look towards the Sun during a total solar eclipse (taking proper precautions of course) and you'll see a beautiful, crown-shaped glow surrounding the Moon. It's the solar corona – a hot plasma that extends millions of kilometres out into space. This region is a million times dimmer than the solar surface beneath, yet, strangely, it's at least 1 million kelvin hotter. Seven decades after the unexpected observation was first made, it's still one of the biggest mysteries in astronomy.

From images carved into ancient stones in places like County Meath, Ireland, it seems that people recorded the appearance of the solar corona at least 5000 years ago. But how do we know anything about the corona given that eclipses occur so rarely – barely a few minutes per year – and then only at specific geographic locations? For what we do know, we can thank the French astronomer Bernard Lyot, who in the 1920s developed a device that can create "artificial eclipses" inside specially adapted telescopes.

By blocking out the extreme brightness of the Sun, this device, known as a coronagraph, allowed astronomers to study the corona for hours, rather than minutes, every year. In fact, in 1943, using data taken during solar eclipses and with coronagraphs at other times, the Swedish physicist Bengt Edlén was able to determine the origin of certain, mysterious spectral lines from the Sun. These, he realized, are emitted when ions of iron and other heavy elements that have been stripped of at least 10 electrons collide with electrons in the corona heated to temperatures of 250 000 K. This estimate was later revised upwards to $10^6$ K and above.

It was a bold claim and some researchers initially struggled to accept the implications because it meant that energy must be flowing from the "cool" 6000 K surface of the Sun into the hotter corona – seemingly in violation of thermodynamics. So began the ongoing, seven-decade search for the non-thermal mechanisms by which energy from the Sun is transported and dissipated to the corona.

Intriguingly, these scientific studies took on a wider political importance during the Second World War. That's because the corona emits highly variable charged particles, magnetic fields, X-rays and extreme ultraviolet (EUV) light, which can cause troublesome variations in the structure of the Earth's ionosphere and trigger "radio blackouts". Keen to ensure the smooth running of military communications, both Axis and Allied scientists began deploying novel coronagraphs to try to anticipate when these blackouts might occur.

After the war, Richard Tousey of the US Naval Research Laboratory even explored UV and X-ray emissions from the Sun using instruments flown on V2 rockets captured from the Germans. But it was not until 1973, when NASA launched its SKYLAB space laboratory – built inside a command module left over from the Apollo lunar missions – that solar EUV and X-ray wavelengths were routinely acquired.

These measurements proved pivotal to our understanding of the corona. As the US astronomer Leo Goldberg from the Kitt Peak National Observatory wrote in the foreword to *A New Sun* – John Eddy's 1979 book of SKYLAB's achievements: "Especially illuminating has been the recognition of the extent to which the Sun's magnetic field is responsible for the structure, dynamics and heating of the Sun's outer layers".

Many space missions since SKYLAB have extended our measurements of the corona beyond EUV and X-ray wavelengths, yielding details of the even-higher-energy gamma radiation it emits. A particularly important role has been played by NASA's Solar Dynamics Observatory satellite, which has been travelling in a geostationary orbit around the Earth since 2010. Its 17 megapixel camera gives us a brand new image of the corona once a second, 24 hours a day, seven days a week.

Today, hundreds of scientists are pondering the vast quantities of data we have of the solar corona. Armed with ever more powerful numerical capabilities, they are seeking to finally understand how the

> With so much information to hand, why are we still arguing over the reason why the corona is so hot?

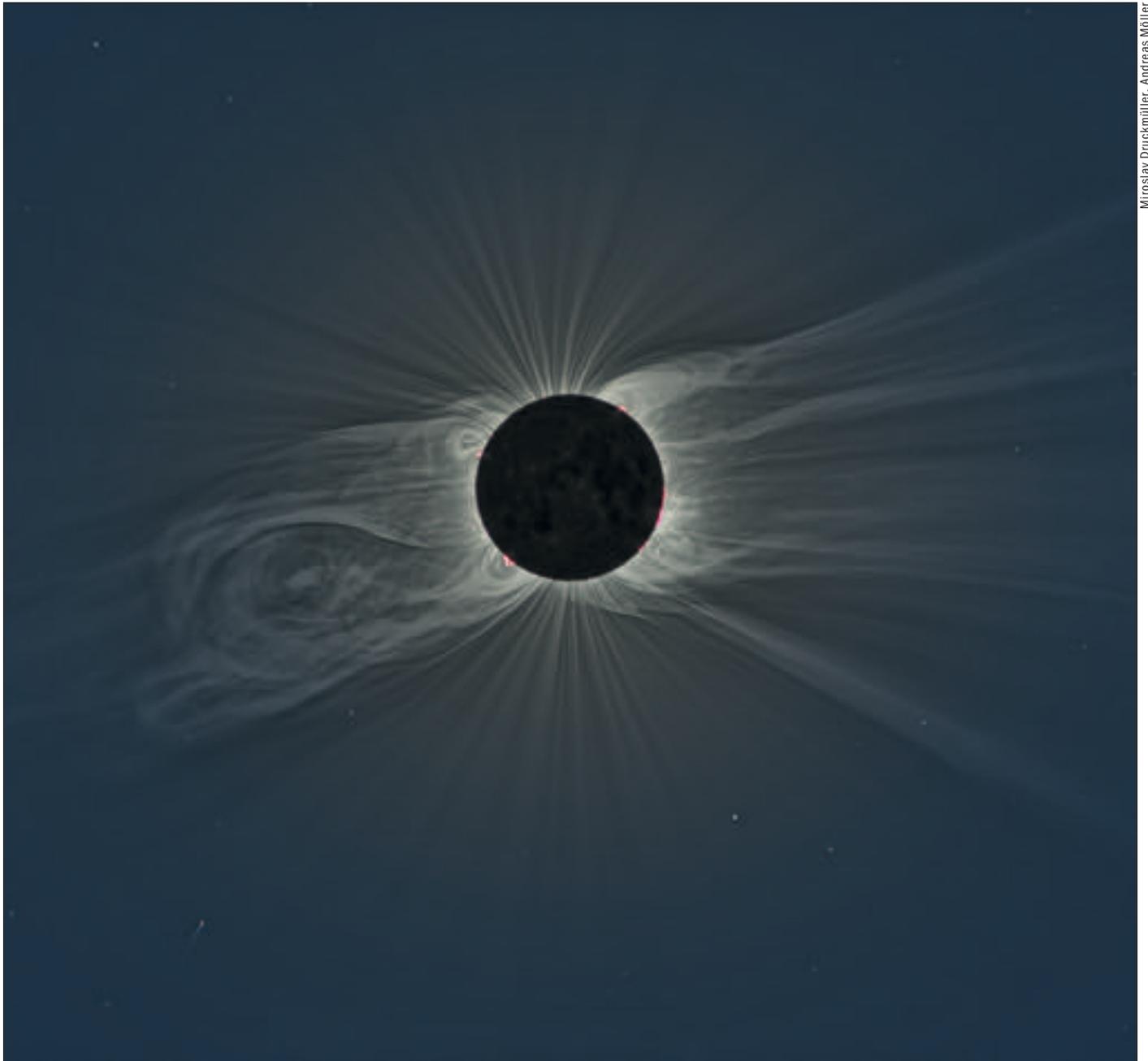

corona and magnetism interact. But with so much information to hand, why are we still arguing over the reason why the corona is so hot? How, in other words, does energy from the heart of the Sun get transported and dissipated above the visible surface?

**An enduring problem**
Ask a group of astronomers what heats the solar corona and you'll be surprised by how many different answers you'll get. Some might say "magnetic energy". Others "Alfven waves" or "nanoflares". Yet others will talk about "turbulence", "ion cyclotron waves" or "magnetic reconnection". In fact, almost one article about the solar corona has been published per day since 1943.

The stage upon which this mystery plays out is straightforward. Nuclear fusion in the heart of the Sun generates energy, of which a tiny fraction (less than 0.001%) ends up as "free magnetic energy". A concept developed in the 19th century by the German physicist Hermann von Helmholtz, free magnetic energy is a reservoir of "ordered" energy that can be converted to less ordered forms. This free energy somehow gets transported and dissipated as heat above the Sun's visible surface. But like catching a burglar emerging red-handed from a window with a bag of loot, we can't easily observe the action and identify the suspected mechanism(s).

In one sense, finding what heats the corona is easy. After all, less than 0.001% of the Sun's entire power is required to sustain the corona and all of it is carried by convective motions just under the surface. But precisely because so little is needed, almost any mode of magnetic-energy transport could sustain the corona, making it tricky to rule out theoretical ideas. As one astronomer joked, "With so many ways to heat the corona, why is it so cold?"

The next challenge arises because the Sun is a

**Sunny times**
The solar corona photographed by Miloslav Druckmüller from Brno University of Technology in the Czech Republic during the 2020 total eclipse over Argentina, showing plasma outflowing to space, bright loops and a coronal mass ejection.

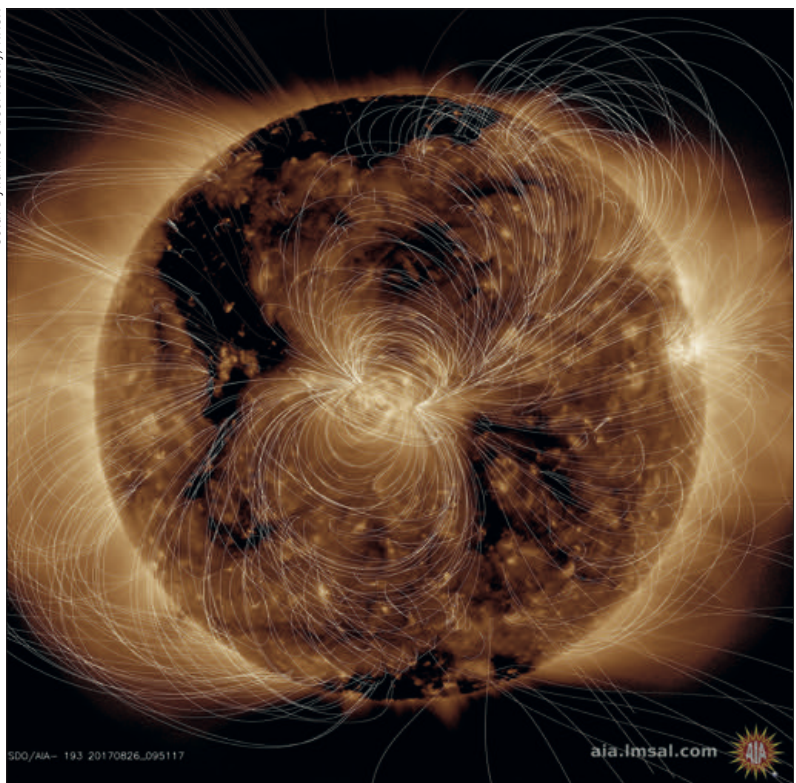

**New views of the Sun** NASA's Solar Dynamics Observatory has been monitoring the Sun's corona since 2010. Shown here is the Sun pictured at extreme ultraviolet wavelengths on 10 August 2018, superimposed with the magnetic-field lines calculated by computer models. The high concentration of field lines (centre) correlates with a bright, active region, while the lack of lines (top) shows a coronal hole, showing that magnetism drives the dynamic activity near the Sun's surface.

plasma: a hot, ionized gas containing free ions and electrons. Like water or any other fluid, the plasma is subject to many nonlinear interactions. Changes in velocity, for example, depend on the velocities themselves, making the flow of the plasma fiendishly difficult to describe. But the corona also has complex nonlinearities arising from the fact that plasma is made up of charged particles.

No large-scale electric fields are sustained in plasma blobs because electrons quickly move to "short out" long-lived electric fields, but magnetic fields can penetrate the corona (there being no "magnetic monopoles" that could short them out). These fields exert a Lorentz force on the moving fluid, altering its bulk velocity and internal electric currents, which in turn modify the magnetic field. Cause and effect become thoroughly confused.

The study of the dynamic behaviour of magnetic fields in an electrically conducting fluid such as a plasma was initiated in 1942 by the Swedish physicist Hannes Alfvén, who dubbed it "magnetohydrodynamics" (MHD). Alfvén went on to win the 1970 Nobel Prize for Physics for his work on MHD, which film enthusiasts might remember as the power behind the fictional submarine-propulsion system in the movie *The Hunt for Red October*. But as Alfvén realized, the freed electrons also lead to non-local effects.

He calculated that in the ideal limit of zero electrical resistance, moving blobs of fluid always belong to a particular set of magnetic lines of force. As they move along those lines, the blobs trace out in time a tube-like structure. But because the blobs can't cross field lines no matter how the fluid moves, the tubes must maintain their topology. Two tubes that aren't interlinked, for example, can't suddenly be made to thread one another.

Alfvén's insight is crucial because the plasma in the solar corona behaves in this almost ideal fashion – except, that is, on very small scales where non-ideal effects involving ions, electrons and their dynamical interactions kick in. In fact, because ideal plasmas have no dissipation, any model for coronal heating must generate dissipation on these tiny scales. In the solar corona, our concept of plasma as a fluid breaks down on the smallest scales.

Simulations suggest that dissipation occurs on a scale of about 100 m at which the constituent ions and electrons can behave separately. On this scale, known as the "ion inertial length", the plasma no longer behaves as a single fluid, as assumed in MHD. To find out what is going on, researchers instead have to solve "kinetic" equations, derived from Ludwig Boltzmann's transport equation, that describe the coupled motions of particles and electric and magnetic fields.

### Calculations and observation

You might wonder why we can't just solve the mystery of the solar corona numerically. Surely cranking through the numbers on a supercomputer can yield answers? It's true that numerical experiments can address various aspects of the problem that can't easily be studied in the lab (a plasma-filled tokamak being used to generate fusion power, for example, will never approach the near-ideal solar-corona conditions). Unfortunately, even the best computers don't have enough memory to tackle the enormous range of scales involved.

To get a sense of the problem, consider the "active regions" on the corona – groups of dark sunspots accompanied by brighter areas that come and go over an 11-year timescale. To capture the transfer of energy via dissipation on the 100 m kinetic scale within a 50 000 km active region, you would need about $10^{17}$ cells, which is impossible with current computers.

This issue is often side-stepped using the qualitative theory of fluid turbulence developed in the mid-20th century by the Soviet physicist Andrey Kolmogorov, in which energy in fluid motions naturally cascades from large to small dissipative scales, following a universal law. Viewed in this way, the problem was "solved" by the Scandinavian physicists Boris Gudiksen and Åke Nordlund in 2005, who used just 3.375 million cells. But is their invocation of Kolmogorov's argument correct? The jury is still out.

You might also wonder whether we can't just take better data. The problem is that our best images of the Sun to date have a resolution of just 200 km. To capture both transport and dissipation in action, we'd need a 45 m-diameter space telescope operating at EUV wavelengths (about 50 nm) coupled to a camera with 400 times more pixels than the largest ever built. That's just not on the cards any time soon.

The story is further complicated by the fact that electrons conduct heat so well. If there happens to be a local burst of heating, they re-distribute that heat over thousands of kilometres to cooler, far-off parts of the corona. It's as if our burglar has quickly covered their tracks, moving evidence far from the original crime scene and disguising it.

The good news is that astronomers have a great track record of using the laws of physics to solve seemingly intractable problems, such as the evolution of elements in our solar system. I am confident that physics can track the problem of the corona too, with much progress having been inspired by an ingenious thought experiment developed in 1972 by the US astrophysicist Eugene Parker from the University of Chicago. Using the MHD equations of motion in the limit of zero resistance, Parker and many other researchers since then have developed a novel picture of the corona.

### Parker's piano

Parker imagined a long, straight volume of plasma entering the corona, with a nearly uniform magnetic field (figure 1a). Now waggle or twist the cylinder from side to side at its base. The magnetic fields that are subject to this motion will exert a tension on the plasma, sending waves along it like a piano wire. The density of the plasma at any location in the cylinder will depend on the precise details of the convection-magnetism interaction that formed it.

According to MHD, these so-called "Alfvén waves" travel more slowly when plasma densities are higher, just as waves on denser piano wires move more sluggishly and have lower notes than on lighter, thinner wires. As these waves move upwards, neighbouring fluid blobs of different densities quickly get out of phase (figure 1b). Their wave energy can then be readily dissipated in a sideways direction by a kind of friction, like bringing two adjacent piano wires too close together.

Over the years, researchers have developed an entire class of models of coronal heating based on this concept. Many believe that this is how magnetic energy – drawn out by the solar wind – gets transported and dissipated into interplanetary space. The notion can even be used to describe strange, dark patches of the corona that were discovered in SKYLAB data. These "coronal holes", which live from weeks to months, can cover as much as 10% of the solar surface.

The solar magnetic fields here can be pictured as cylinders bent smoothly so that both ends are attached to the heavier, surface plasma (figure 1c). To make calculations of coronal heating easier, however, Parker imagined straightening out the cylinders again, like piano wires fixed at both ends (figure 1d). Waves sent from one end of the surface reflect off the other end, with the tube amplifying frequencies that match the appropriate natural frequency (just as piano wires amplify waves matching their natural frequency).

Such a picture (figure 1e) was developed mathematically in 1978 by the physicist James Ionson,

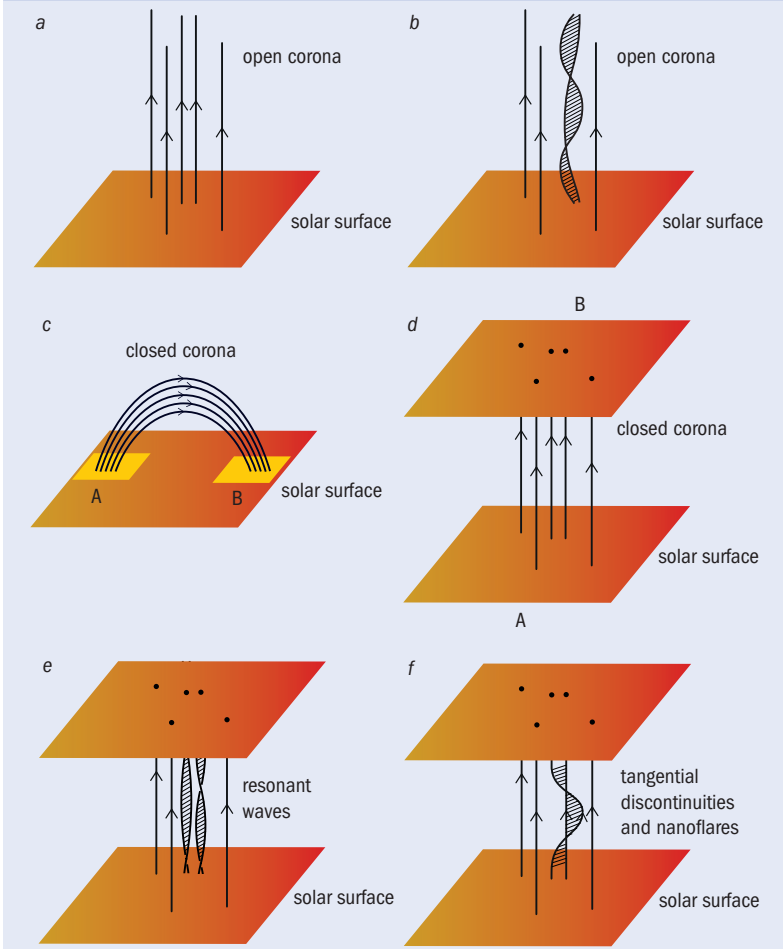

**1 Understanding the corona**

Inspired by ideas from the US astronomer Eugene Parker, these diagrams show how large-scale motions within the solar corona can dissipate energy at small scales (shaded regions). Imagine here a straight set of "open" magnetic field lines emerging from the solar surface (a) being subject to sideways motion (b), with neighbouring parts of the corona with different densities quickly getting out of phase, dissipating energy sideways at small scales by a kind of friction. In the case of field lines (c) that double-back to the solar surface – so-called "plasma loops" – these are straightened out in Parker's model, like piano wires attached and fixed at both ends (d). When rapid motions across the surface A induce resonances in the wave motions, then small scales (shaded areas) can develop owing to resonant absorption (e) or by so-called "tangential discontinuities" forming at the tangled surfaces between more slowly moving field lines (f). In the latter case, this gradually builds up energy and leads to the creation of "nano flares". With a billionth of the energy of ordinary solar flares, these flares are currently being hunted down by astronomers.

who later rose to fame as head of research for Ronald Reagan's proposed "Star Wars" missile-defence initiative. But can this MHD approach tell us what is happening in the corona at scales of 100 m or less? Apparently so, according to Peter Goldreich of Caltech and Seshadri Sridhar of the University of Toronto, who in the mid-1990s showed how oppositely directed Alfvén waves can lead to a turbulent cascade to small scales.

Parker noted, however, that most of the power in observable surface motions occurs on timescales of minutes, not tens of seconds as was needed in Ionson's model. He therefore wondered what structure you get if these slower motions waggle and/or twist one end of the tube, while the other is kept fixed. The answer turned out to be as important as it was

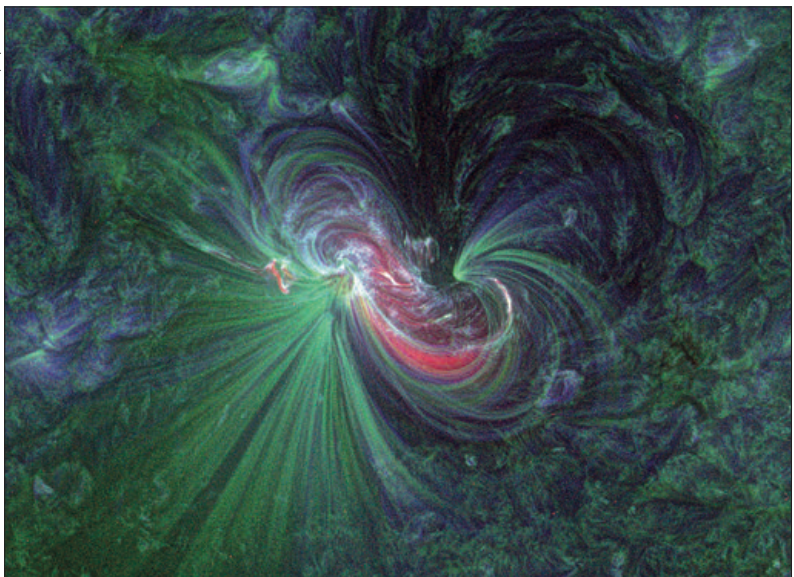

**Colourful corona** This image shows a part of the solar corona observed over an active region using the Atmospheric Imaging Assembly (AIA) on NASA's Solar Dynamics Observatory on 12 July 2017. It is a composite of extreme-ultraviolet light emitted by highly ionized atoms in the corona at temperatures of 5 million kelvin (red), 1–2 million kelvin (blue) and 0.8 million kelvin (green). The various structures are created by coronal plasma trapped by the Sun's magnetic fields. The origin of such high coronal temperatures is still a mystery.

> What we hope to find are some fingerprints, some smoking gun or some statistical patterns from observations to reveal the heating mechanisms unexpected: unless surface motions are unnaturally well-ordered, you get sudden changes in direction between neighbouring magnetic tubes of plasma (figure 1*f*).

Known as "tangential discontinuities" (or TDs), these are sheets of electrical current that store free energy on small scales. As the sheets inevitably get thinner, the currents can be so big that instabilities and kinetic effects end up dissipating the slowly built-up energy even without having to invoke turbulence. Like a Shakespearean tragedy, the apparent purity of the theory leads to its own demise.

**The hunt for nanoflares**
Parker argued that the closed solar corona should release these currents in small bursts of energy, dubbed "nanoflares", which have an energy of about $10^{16}$ J – roughly a billionth of a typical, large flare. Astronomers have been on the hunt for nanoflares ever since he predicted them in 1988 and every so often an article pops up claiming to have detected nanoflares – thus solving the heating problem. But such studies are at the limit of what we can detect, and they may be just a manifestation of (unobservable) smaller-scale structures associated with entirely different mechanisms.

Ultimately, the solution to the coronal-heating problem must come from observations of the Sun. But without accurate measurements of coronal magnetic fields, we cannot trace the energy flow and even massive solar flares leave behind only small changes in the magnetic field on the solar surface. Observers are left to examine the scene of a crime that has been confused and thoroughly cleaned up by heat conduction.

What we hope to find are some fingerprints, some smoking gun or some statistical patterns from obser- vations to reveal the heating mechanisms. Unfortunately, nothing definite has so far appeared. A review published by Cristina Mandrini from the University of Buenos Aires and colleagues in 2000 listed an embarrassing 22 different models, none of which have yet been eliminated. To use the immortal words of Wolfgang Pauli, our models are "not even wrong".

Still, it seems we have a basic understanding of coronal heating in both open and closed (loop) structures. But what about some of the other proposed suggestions for the solar-corona problem? One includes "ion cyclotron waves" – Alfvén waves at kilohertz frequencies in resonance with the helical motion of ions. Another is "magnetic reconnection" – whereby plasmas change their topology, allowing the magnetic field lines to diffuse from their original plasma, often leading to dramatic consequences such as solar flares and coronal mass ejections.

Magnetic reconnection is often dynamic and self-sustaining, but in MHD it cannot by itself lead to much heating, instead generating kinetic energy in outward flows of plasma as the magnetic fields seek a new equilibrium. It may, however, subsequently lead to significant plasma heating and is believed to power enormous flares. Indeed, numerical experiments have revealed that these reconnection processes can generate tiny, balls of plasma-containing magnetic bubbles that enhance local plasma heating. The "plasmoids" may, in turn, generate ion-cyclotron waves at kilohertz frequencies, which could heat the plasma, according to recent analyses of coronal spectral lines.

**Final frontier**
It's clear that magneto-hydrodynamics can successfully describe the propagation, storage and development of small scales needed to dissipate free magnetic energy. However, we will need physics beyond this approximation to determine exactly how the magnetic energy gets converted into the random motions of particles in the plasma.

Coronal heating is a challenging yet fascinating topic that, I hope, future generations of researchers will be inspired to explore. And with lots of data coming in from NASA's Parker Solar Probe and Europe's Solar Obiter – as well as the Daniel K Inouye Solar Telescope in Hawaii – there will certainly be lots to keep people busy for many years to come. ∎